\documentclass[nonacm=true, sigplan]{acmart}
\renewcommand\footnotetextcopyrightpermission[1]{} 
\usepackage[utf8]{inputenc}
\usepackage[T1]{fontenc}
\pdfoutput=1
\usepackage{listings}
\usepackage{mathpartir}
\usepackage{framed}
\usepackage{amsmath}
\usepackage{caption}
\newcommand{\return}[0]{\\~\\}

\usepackage[
  smaller,
  printonlyused,
  nolist
]{acronym}

\definecolor{light-gray}{rgb}{0.75,0.75,0.75}
\definecolor{dark-gray}{rgb}{0.55,0.55,0.55}
\definecolor{key-color}{rgb}{0.5,0,0.5}
\lstdefinestyle{JastAdd}{
  language=Java,
  frame=none,
  showstringspaces=false,
  columns=flexible,
  basicstyle={\small\ttfamily},
  numbers=none,
  keywordstyle={\bfseries},
  otherkeywords={syn,inh,aspect,coll,each,contributes,when,rewrite,eq,to},
  deletekeywords ={int},
  breaklines=true,
  breakatwhitespace=true,
  tabsize=2,
  captionpos=b,
  mathescape=true
}

\lstdefinestyle{SPL}{
  language=C,
  frame=none,
  showstringspaces=false,
  columns=flexible,
  basicstyle={\small\ttfamily},
  numbers=none,
  keywordstyle={\bfseries},
  otherkeywords={proc, var, val, while, if, else, ref},
  deletekeywords ={int},
  breaklines=true,
  breakatwhitespace=true,
  tabsize=2,
  captionpos=b,
  mathescape=true
}

\begin{document}
\title{Patterns for Name Analysis and Type Analysis with JastAdd}
\author{Uwe Meyer}
\email{{uwe.meyer, bjoern.pfarr}@mni.thm.de}
\author{Björn Pfarr}
\affiliation{%
  \institution{Technische Hochschule Mittelhessen / 
University of Applied Sciences}
  \department{Dept. of Mathematics, Natural Sciences, Computer Science}
  \streetaddress{Wiesenstrasse 10}
  \city{Giessen}
  \country{Germany}}

\begin{abstract}
In the last two decades, tools have been implemented to more formally specify the semantic analysis phase of a compiler instead of relying on handwritten code.
In this paper, we introduce patterns and a method to translate a formal definition of a language's type system into a specification for JastAdd, which is one of the aforementioned tools based on \acfp{rag}. This methodological approach will help language designers and compiler engineers to more systematically use such tools for semantic analysis. 
As an example, we use a simple, yet complete imperative language and provide an outlook on how the method can be extended to cover further language constructs or even type inference. 
\end{abstract}

\maketitle
\section{Introduction}
Compiler construction is one of the oldest and most mature disciplines in computer science. Although some major results such as Chomsky's work on formal languages and grammars \cite{Chomsky:1956} or Knuth's work on \acf{ag} \cite{Knuth68} are now more than 50 years old, there is one aspect of compiler construction that seems to be a focus of research only since the last two decades. Whilst, mostly due to the groundbreaking work mentioned above and others, lexical analysis and syntax analysis are very well understood and available thanks to tools as Lex and Yacc \cite{Johnson1975YaccYA} (as well as their successors), ANTLR \cite{ANTLR} and PEGs \cite{Ford:2004:PEG:982962.964011}, tool support for type analysis has been very limited. This leads to the fact that, even in teaching, these parts are usually handcrafted in many lines of code using techniques such as the visitor pattern \cite{gamma1994design} \cite{books/cu/Appel1998} to traverse the abstract syntax tree and apply code fragments to each node type.

Nowadays, there exist few tools to generate the code for the later phases, such as JastAdd \cite{Ekman2007}, Kiama \cite{conf/gttse/Sloane09} and Silver\cite{journals/entcs/WykBGK08}.
There are several successful language implementations using these tools and although they are mainly from the groups who have developed the respective tools, it seems that translating a language's semantics into the specification languages of these tools is very much a major design effort.

This paper aims at providing a systematic approach on how to translate a language's formal type system into rules for JastAdd, a framework developed by G. Hedin and her group at Lund University \cite{Ekman2007} based on her previous work on \acp{rag}\cite{Hedin99}.
We will formally describe the type system of a simple imperative language using Cardelli's notation \cite{Cardelli97} and show how JastAdd rules can be derived from it.
Hedin\cite{Hedin99} and members of her group provide hints on how to use JastAdd for example for PicoJava, a subset of Java, but we would like to show how a  type system can be transcribed into JastAdd code in a more formal manner.

\subsection{The \acf{spl}}
The \acf{spl} \cite{SPL} has been developed as a language for a 2nd-year compiler construction course at
\begin{anonsuppress}
 our 
\end{anonsuppress}
university. Besides learning the theory of automata, formal languages and compiler construction itself, students have to implement a complete compiler for this language from lexical analysis to generation of assembler code within one term. While flex\cite{Paxson:1988:FFL}, JFlex\cite{JFlex}, Bison\cite{Donnelly:1988:BYC}, and Cup \cite{CUP} are used for the lexical and syntactic analysis, the remaining phases have to be implemented by hand in Java or C.

Needless to say, in order to be simple, \ac{spl} provides only a handful of procedural language constructs. Among them are assignments, loops, branches and procedures. Arguments are passed either by call-by-value or call-by-reference. There are primarily two different types available to a user, namely an array type constructor with a static length and the primitive type int. Although boolean values and the boolean type are internally used for conditions, they cannot be accessed by the user in a direct way (e.g. declaring a variable of a boolean type is not allowed). In general, types are compared following reference semantics. 
That is, the reference of the internal type graph is compared whenever two types need to be checked for their equivalence. 
Because each use of the array type constructor yields a new type and therefore is inequal to any other type, type synonyms allow to introduce aliases by referencing existing types.

\ac{spl}'s formal type system is defined using the abstract syntax that is shown in figure \ref{figure:abstract-syntax}. 
The notation $\overline{X}$ is an abbreviation for $X_1, ..., X_n$.

\begin{figure}[H]
\begin{alignat}{3}
  Identifiers & \quad x \quad && \quad & & \quad \nonumber \\
  Numerals & \quad \ell \quad && \quad & & \quad \nonumber \\
  Program & \quad P && ::= & & \quad program \; \overline{D} \nonumber \\
  Types & \quad \tau && ::= & & \quad x \nonumber \\
    \quad & \quad && \quad | & & \quad int \nonumber \\
    \quad & \quad && \quad | & & \quad array[\ell] \; of \; \tau \nonumber \\
  Parameters & \quad \rho && ::= & & \quad ref \; x: \tau \nonumber \\
    \quad & \quad && \quad | & & \quad val \; x: \tau \nonumber \\ 
  Variables & \quad V && ::= & & \quad var \; x: \tau \; \nonumber \\
  Declarations & \quad D && ::= & & \quad proc \; x(\overline{\rho})\;\{\;\overline{V}\;S\;\} \nonumber \\
    \quad & \quad && \quad | & & \quad type \; x = \tau \nonumber \\
  Statements & \quad S && ::= & & \quad T_i := T_j \nonumber \\
    \quad & \quad && \quad | & & \quad x(\overline{T}) \nonumber \\
    \quad & \quad && \quad | & & \quad \overline{S} \nonumber \\
    \quad & \quad && \quad | & & \quad if \; (T) \; S_i \; else \; S_j \nonumber \\
    \quad & \quad && \quad | & & \quad while \; (T) \; S \nonumber \\
    \quad & \quad && \quad | & & \quad skip \nonumber \\
  Terms & \quad T && ::= & & \quad x \nonumber \\
    \quad & \quad && \quad | & & \quad T_i[T_j] \nonumber \\
    \quad & \quad && \quad | & & \quad \ell \nonumber \\
    \quad & \quad && \quad | & & \quad T_i +_{bin} T_j \nonumber \\
    \quad & \quad && \quad | & & \quad T_i \leq_{cmp} T_j \nonumber \\
    \quad & \quad && \quad | & & \quad -T \nonumber
\end{alignat}
\caption{The abstract syntax of \ac{spl}}
\label{figure:abstract-syntax}
\end{figure}

Listing \ref{figure:spl-sample} shows an exemplary \ac{spl} program that calls a procedure \texttt{gcd} which computes the greatest common divisor of two numbers and returns the result using a third, reference parameter. Except for the \texttt{\#}-symbol, which stands for the inequality operator, the semantics of \ac{spl} is straightforward and should not pose any hurdle to a reader familiar with typical procedural programming languages.

\begin{minipage}{\linewidth}
\begin{lstlisting}[
  style=SPL,
  caption={An exemplary \ac{spl} program},
  label=figure:spl-sample
]
proc main() {
  var result: int;
  gcd(25, 15, result);
  printi(result);
}

proc gcd(val a : int, val b: int, ref result: int) {
  while (a # b) {
    if (a < b) {
      b := b - a;
    } else {
      a := a - b;
    }
  }
  result := a;
}
\end{lstlisting}
\end{minipage}

\subsection{JastAdd}

JastAdd was initially developed by Hedin in 1999 as the result of her research on \acf{rag} \cite{Hedin99}. Many extensions have been proposed and implemented since then. 

Its core principle can be best described as aspect-oriented \acp{rag}: Given an abstract syntax specification, attributes are defined by equations assembled in aspect files.

The abstract syntax is specified in form of an extension to \acp{rag}, namely object-oriented \acp{rag} \cite{Hedin1989AnON}. That is, the underlying \ac{cfg} is interpreted as a hierarchy of classes. While the non-terminal on the left-hand side of a production rule leads to an equally named class, the symbols on the right-hand side denote members of this class. 
The types of these members as well as inheritance of the class can be expressed using annotations. 
Most interestingly, attributes can be interpreted as methods of this class, which ultimately led to the introduction of parameterized attributes \cite{Hedin2000ReferenceAG}. 
Because of this correlation between class hierarchies and \acp{cfg}, we sometimes refer to non-terminals as nodes (in terms of the nodes of an \ac{ast} or derivation tree) or classes.

Besides the object-oriented view on \acp{rag} there are different extensions that relate more to the actual attributes. 
In the context of this paper the most notable of these extensions are:
\begin{itemize}
  \item \textbf{Broadcasting} automatically propagates the value of an inherited attribute down to all child nodes. This feature is especially useful when many simple copy equations would be necessary to accomplish the same otherwise.
  \item \textbf{Collection Attributes} denote a feature that allows the contribution of values to attributes that are marked as such and furthermore may be arbitrary far away in the \ac{ast}. In some sense this feature is the synthesizing counterpart of broadcasting, since it allows to propagate values up the \ac{ast} by ``sending'' them directly to the destination attribute.
  \item \textbf{Rewriting}, as the name suggests, allows to rewrite nodes or even whole subtrees under a selected condition.
\end{itemize}

Besides those extensions, JastAdd still offers the basic concept of synthesized and inherited attributes just like \acp{ag}.

\section{Name Analysis}
\subsection{Approach}
As basis for our investigations, we firstly developed a compiler using Java with JFlex and CUP for the lexical and syntactical analysis. The remaining phases (i.e. name analysis, type analysis and code synthesis) were implemented by hand. Next, we reused the existing code of the lexical and syntactical analysis to implement another compiler using JastAdd as a framework.
While the former compiler comprises approximately 2100 lines of code (excluding the scanner and parser), the latter consists of approximately only 700 lines of JastAdd specification files. 

It turned out that development using JastAdd is more efficient than conventional approaches like our first implementation, as it is more abstract and declarative. Performance in terms of space and time for required for compilation of larger SPL programs is slightly better for the handcrafted compiler.

After incrementally generalizing the attributes used in the JastAdd implementation, a pattern emerged in the name and type analysis that eventually led to the methods  and patterns  described in the following.

\subsection{Free Variables}\label{section:name-analysis}
We will now start examining the name analysis phase of a compiler \cite{dragonBook}, which collects information about all names occurring in the source program and maps it to semantic information such as the kind (variable, type, procedure, ...) and detailed attributes such as the type graph. All this information is stored in a symbol table, which usually consists of multiple levels according to the scoping rules of the language.

For example, \ac{spl} defines two different scoping levels: One for local declarations such as variables and parameters and one for global declarations like type synonyms and procedures. These scoping rules can be formalized by specifying the set of free \textit{variables}, which ultimately requires to specify how \textit{variables} are bound by declarations \cite{Cardelli97}. In this context \textit{variables} are not to be confused with local variables of \ac{spl}. While the former is formally used to denote all identifiers (i.e. even those that are associated with procedures or types, for example), the latter denotes only identifiers that are bound by the variable declaration of \acp{spl}. For disambiguation, we write the former in italics. Figure \ref{figure:binders} and \ref{figure:fv-declarations} show the relevant parts of \ac{spl}'s specification for \textit{binders} and free \textit{variables}.

\begin{figure}[H]
\begin{alignat}{1}
Parameters \quad & binders(ref\;x:\;\tau) = \{ x \} \nonumber \\
& binders(val\;x:\;\tau) = \{ x \} \nonumber \\
Variables \quad & binders(var\;x:\;\tau) = \{ x \} \nonumber \\
Declarations \quad & binders(type\;x = \tau) = \{ x \} \nonumber \\
& binders(proc\;x(\overline{\rho})\;\{\;\overline{V}\;S\;\}) = \{ x \} \nonumber \\
Sequence \quad & binders(D_1,\;D_2,\;...,\;D_n) = \bigcup_{i = 1}^n binders(D_i) \nonumber
\end{alignat}
\caption{The sets of \textit{binders} for declarations}
\label{figure:binders}
\end{figure}

\begin{figure}[H]
\begin{alignat}{1}
  Parameters \quad & fv(ref\;x:\;\tau) = fv(\tau) \nonumber \\
  & fv(val\;x:\;\tau) = fv(\tau) \nonumber \\
  Variables \quad & fv(var\;x:\;\tau) = fv(\tau) \nonumber \\
  Declarations \quad & fv(type\;x = \tau) = fv(\tau)  \nonumber \\
  & fv(proc\;x(\overline{\rho})\;\{\;\overline{V}\;S\;\}) = fv(\overline{\rho}) \cup fv(\overline{V}) \nonumber \\
  & \quad \cup (fv(S) - (binders(\overline{\rho}) \cup binders(\overline{V}))) \nonumber \\
  Program \quad & fv(program \; \overline{D}) = fv(\overline{D}) - binders(\overline{D}) \nonumber \\
  Sequence \quad & fv(D_1,\;D_2,\;...,\;D_n) = \bigcup_{i = 1}^n fv(D_i) \nonumber
\end{alignat}
\caption{The sets of free \textit{variables} for declarations}
\label{figure:fv-declarations}
\end{figure}

The free \textit{variables} of types, statements and expressions are defined in a straightforward manner, where each occurrence of an identifier $x$ results in a free \textit{variable} $x$.

\subsection{A Method for Deriving a \texttt{lookup}-Attribute}
\label{section:a-method-for-lookup}

Given a formal definition of free \textit{variables} such as the one above, it is possible to derive the rules that are necessary to create and fill a symbol table during the name analysis of a conventional compiler. However, by taking advantage of the reference semantics of \acp{rag}, an explicit symbol table can be avoided. Instead, a parameterized attribute can be used to associate names with a reference to the \ac{ast} node of the corresponding declaration.

The general method for the introduction of such a parameterized attribute (which we call \texttt{lookup}) is as follows:
\begin{enumerate}
\item Introduce a new inherited attribute \texttt{lookup} to all nodes of the \ac{ast}.
\item For each language construct that defines a scope,
  \begin{enumerate}
    \item define a new attribute \texttt{declarations} that holds a list of all of its containing declarations.
		\item define a new attribute \texttt{lookupInScope} (e.g. \texttt{lookup\-Global}, \texttt{lookupProcedure}, \texttt{lookupBlock}) which\\searches for the name using the attribute \\\texttt{declarations}.
		\item add an equation for the \texttt{lookup} attribute of its direct children in the \ac{ast} to override the searching behavior.
  \end{enumerate}
\end{enumerate}

Besides inherited and parameterized attributes, this proceeding requires broadcasting to unfold its full potential. Broadcasting enables a programmer to define the value of an inherited attribute at an appropriate location in the \ac{rag}. JastAdd then takes care of propagating this value to the non-terminal for which the attribute was actually defined.

The language constructs that define a scope can easily be read off from the definition of $fv$. As shown in figure \ref{figure:fv-declarations} there are two equations that use the \textit{binders} function to bind free occurrences of \textit{variables}. It is these equations which indicate language constructs that define scopes. Consequently, step 2 of the procedure above needs to be applied to procedures and programs.

Moreover, it can be read off which scope binds which of the free \textit{variables}. For example, in \ac{spl} a procedure's local variables and parameters bind any of the corresponding free occurrences in the procedure's statement. On the contrary, the free \textit{variables} that may occur in the types of local variable and parameter declarations are not bound by them (as it is implied by the mathemtical precedence). That is, free occurrences of procedures and types are bound on the global level, which can be observed in the equation for programs. All in all, this observation indicates the set of declarations that need to be collected in step 2a of our method.

\subsubsection{An Exemplary \texttt{lookup}-Attribute for \ac{spl}}

To demonstrate and further explain the presented method we will use \ac{spl}'s definition of $fv$ and $binders$ to introduce a \texttt{lookup} attribute.

Listing \ref{figure:inherited-lookup} shows the inherited \texttt{lookup}-attribute as it should be introduced by step 1 of the method. As context information is usually distributed across the syntax tree, it must be passed down to the relevant positions (e.g. a procedure call) from the nearest parent that is able to fully collect this information (e.g. the program). This is the reason why the \texttt{lookup}-attribute must be an inherited one:

\vspace*{3mm}
\begin{minipage}{\linewidth}
\begin{lstlisting}[
  style=JastAdd,
  caption=The inherited \texttt{lookup}-attribute,
  label=figure:inherited-lookup
]
inh Optional<Declaration> Tree.lookup(String name);
\end{lstlisting}
\end{minipage}

Since a programmer may use a variable that is not declared, the \texttt{lookup}-attribute returns an \texttt{Optional} in order to be total. Overall, an attribute declaration as it can be seen in listing \ref{figure:inherited-lookup} can be read in a manner similar to a Java method signature. In fact, parameterized attributes can even be translated to virtual functions \cite{Swierstra91} \cite{Hedin2013}.
\return
The second part of the method requires more effort and may vary depending on the actual source language. According to step 2a, we first need a new attribute \texttt{declarations} for each language construct that defines a scope. As previously mentioned, those language constructs can be read off from the definition of $fv$. In case of our simple source language \ac{spl} there are exactly two language constructs that define a scope: programs and procedures.

Following step 2a, the \texttt{declarations}-attribute should contain all declarations that are part of such a language construct. In case of a program this attribute is equal to \texttt{getDeclarationList} (see listing \ref{figure:declarations-attribute}). However, in case of a procedure this attribute is the union of \texttt{getParameterList} and \texttt{getVariableList} (see listing \ref{figure:declarations-attribute}). Unsurprisingly, these lists can be read off from the right-hand sides of the definition of $fv$ for programs and procedures, as they correspond to the sets of binders that are used to capture the free variables of the respective scope.

\begin{minipage}{\linewidth}
\begin{lstlisting}[
  style=JastAdd,
  caption={The \texttt{declarations}-attribute for programs and procedures},
  label=figure:declarations-attribute
]
syn Collection<Declaration> Program.declarations() {
  ArrayList result = new ArrayList<>();
  getDeclarationList().forEach(result::add);
  return result;
}

syn Collection<Declaration> Procedure
  .declarations() {
    ArrayList result = new ArrayList<>();
    getParameterList().forEach(result::add);
    getVariableList().forEach(result::add);
    return result;
  }
\end{lstlisting}
\end{minipage}

For reasons of error handling (see subsection \ref{section:dealing-with-naming-errors}), it is necessary that the \texttt{declarations}-attribute is not implemented using a set in the mathematical sense but instead any kind of collection which is able to hold duplicate elements.
\return
According to the next step 2b, we need an additional new attribute \texttt{lookupInScope} for programs and procedures. Since those two language constructs correspond to the global and local declaration level, we named their \texttt{lookupInScope}-attributes \texttt{lookupGlobal} and \texttt{lookupLocal}
in the listing \ref{figure:lookup-in-scope}.

\begin{minipage}{\linewidth}
\begin{lstlisting}[
  style=JastAdd,
  caption=The \texttt{lookupGlobal}- and \texttt{lookupLocal}-attribute for programs,
  label=figure:lookup-in-scope
]
syn Optional<Declaration> Program
  .lookupGlobal(String name) = declarations()
    .stream().filter(d -> d.getName().equals(name))
    .findFirst();

syn Optional<Declaration> Procedure
  .localLookup(String name) = declarations()
    .stream().filter(d -> d.getName().equals(name))
    .findFirst();
\end{lstlisting}
\end{minipage}

As it can be very easily seen, both equations in listing \ref{figure:lookup-in-scope} look the same. Unfortunately, JastAdd does not allow to define multiple attributes by one equation. That is, unless there is a common superclass that identifies these language constructs as scopes, there is currently no way to remove this redundancy in JastAdd.
\return
The last step 2c ensures that JastAdd consolidates the scope information such that it can be used everywhere. It completes the initial declaration of the inherited \texttt{lookup}-attribute of the first step. As mentioned previously (see section \ref{section:a-method-for-lookup}), the broadcasting feature of JastAdd plays an important role in this step. Usually, i.e. without this feature, it would be necessary to add an equation for all subclasses of \texttt{Tree} to properly complete the \texttt{lookup}-attribute. However, many of these equations would be trivial in that they would just pass the value on to the direct children. With broadcasting support, this is handled automatically by the tool. Consequently, it is sufficient to define an inherited attribute by providing a single equation for it at the root node of the \ac{ast}. In JastAdd this even works if the root node is not a direct predecessor of the node for which the attribute was defined.

Back to step 2c of our method, this means that it is sufficient to add an equation for the direct descendants of procedures and programs. This overwrites the default behavior of the broadcasting feature, which would simply copy the information of the parent's scope.

As implied by the first three equations in listing \ref{figure:spl-lookup}, the direct descendants of a \texttt{Procedure} are \texttt{getStatement}, 
\newline
\texttt{getVariable(int index)} and \texttt{getParameter(int index)}. The latter two notations allow the programmer to define an equation for each element of \texttt{getVariableList} and \texttt{get\-ParameterList}, respectively.

The only direct descendant of a \texttt{Program} is its list of declarations, which is why there is only one equation in this case. What is apparent when comparing the equations of \texttt{Procedure} and \texttt{Program}, is that the former does not only use the \texttt{lookupInScope}-attribute but also the \texttt{lookup}-attribute itself. This is due to the usual hierarchy of lexical scopes, as it is necessary to search for a declaration in an upper scope if it is undefined in the current one. Little attention has to be paid regarding the context of the Java expressions: The \texttt{lookup}-attribute on the right-hand side refers to that of the \texttt{Procedure} and not of its descendants, which means that it, in fact, searches in the upper scope.

\begin{lstlisting}[
  style=JastAdd,
  caption=The inherited general look-up attribute,
  label=figure:spl-lookup
]
eq Procedure.getStatement()
  .lookup(String name) = localLookup(name)
    .map(Optional::of)
    .orElseGet(() ->$\;\;$lookup(name));

eq Procedure.getVariable(int index)
  .lookup(String name) = localLookup(name)
    .map(Optional::of)
    .orElseGet(() ->$\;\;$lookup(name));

eq Procedure.getParameter(int index)
  .lookup(String name) = localLookup(name)
    .map(Optional::of)
    .orElseGet(() ->$\;\;$lookup(name));

eq Program.getDeclaration(int index)
  .lookup(String name) = globalLookup(name);
\end{lstlisting}

With the four presented listings \ref{figure:inherited-lookup}, \ref{figure:declarations-attribute}, \ref{figure:lookup-in-scope} and \ref{figure:spl-lookup} the implementation of the scoping rules is already complete. Of course, they can be extended freely to provide further features such as predefined standard declarations. In the context of our compiler implementation we were able to add predefined declarations using higher-order \acp{ag} (also called \acfp{nta} by JastAdd) and by appropriately extending the \texttt{declarations}-attribute of programs.

\subsection{Dealing with Naming Errors}
\label{section:dealing-with-naming-errors}

Up to this point, we ignored the fact that a program may be erroneous in that a programmer may have declared a name twice or more, for example. In this case, the \texttt{lookup}-attribute would simply return one of those conflicting declarations, not indicating that there is something wrong. The other way around, if a name is used that was not previously declared by the programmer, the \texttt{lookup}-attribute just returns \texttt{Optional.empty()}.

Of course, it is highly desired to find all these errors during the name analysis. Following the idea of Boyland \cite{boyland:96phd}, 
collection attributes can be used to specify the respective error contributions in a precise way. However, it is not as easy as with our previous method and in some cases even not possible at all, to derive the cases of errors just by inspecting the definition of \textit{fv}. 
Consequently, in addition to the formal definition of \textit{fv} and \textit{binders}, two more constraints are required to complete the specification for the scoping rules. 

\vspace*{2mm}
\begin{minipage}{\linewidth}
\begin{lstlisting}[
  style=JastAdd,
  caption={Collection attributes}, 
  label=figure:name-errors
]
coll ArrayList<String> Program.nameErrors();
\end{lstlisting}
\end{minipage}

The definition of the collection attribute which contains all errors is given in listing \ref{figure:name-errors}. 

As it can be seen there, this notation resembles that of synthesized attributes. The result type can be chosen quite freely. However, there has to be a default constructor and a method \texttt{add} that accepts a new element. Both requirements apply to Java's standard \texttt{ArrayList}. The element type of the collection (which is \texttt{String} here) determines the type of the contributions we will see later.

\subsubsection{Use of Undeclared Variables}
\label{section:undeclared-variables}

In \ac{spl} as in many other programming languages it is not allowed to use a $variable$ which was not previously declared. During type analysis this means that looking up a name yields no result (e.g. \texttt{Optional.empty()} in our case). Formally, this restriction can be expressed by enforcing that correct programs may not have any free $variables$, which is shown in figure \ref{figure:no-free-variables}.

\begin{figure}[H]
\begin{alignat}{1}
\forall \; P : fv(P) \neq \emptyset \Rightarrow P\;is\;erroneous \nonumber
\end{alignat}
\caption{A program may not have any free $variables$}
\label{figure:no-free-variables}
\end{figure}

Given such a restriction and a way to check the premise, it is rather easy to derive an error contribution from it. Transferred to our simple source language \ac{spl}, for example, the premise can be checked by testing whether the result of the \texttt{lookup}-attribute is present or not for all \textit{variables} in the program. This leads us to the error contribution as it is shown in listing \ref{figure:undeclared-variables-contribution}.

\vspace*{3mm}
\begin{minipage}{\linewidth}
\begin{lstlisting}[
  style=JastAdd,
  caption={Error contribution for undeclared \textit{variables}}, 
  label=figure:undeclared-variables-contribution
]
Identifier contributes "undefined variable"
  when !lookup(getName()).isPresent()
  to Program.nameErrors();
\end{lstlisting}
\end{minipage}
\vspace*{3mm}

Given that any use of a \textit{variable} is represented by an instance of the non-terminal \texttt{Identifier}, it is then sufficient to check the collection attribute \texttt{nameErrors} in order to verify the absence of undeclared \textit{variables} in a program.

\subsubsection{Duplicate Declarations}
\label{section:duplicate-declarations}

Similar to the restriction that undeclared variables are prohibited, it is usually also disallowed to define the same $variable$ twice in the same scope. Some simple languages may avoid conflicts of this kind, e.g. when there are only language constructs that allow to introduce one $variable$ per scope at a time. However, in \ac{spl} this is not the case, which is why the additional restriction in figure \ref{figure:no-duplicate-variables} is required to complete the specification of the scoping rules.

\begin{figure}[H]
\begin{alignat}{1}
\forall \; program\;\overline{D} &: (\bigcap_{i=1}^n binders(D_i)) \neq \emptyset \nonumber \\
  &\Rightarrow program\;\overline{D} \; is \; erroneous \label{figure:no-duplicate-variables:1} \\
\forall \; proc\;x(\overline{\rho})\;\{\;\overline{V}\;S\;\} &: (\bigcap_{D\;\in\;(\overline{\rho}\;\cup\;\overline{V})} binders(D)) \neq \emptyset \nonumber \\
  &\Rightarrow proc\;x(\overline{\rho})\;\{\;\overline{V}\;S\;\} \; is \; erroneous \label{figure:no-duplicate-variables:2}
\end{alignat}
\caption{Programs and procedures may not have duplicate $variables$}
\label{figure:no-duplicate-variables}
\end{figure}

Again, given these two restrictions and a way to check for their premises, it is easy to derive error contributions for them. As the premise verifies that there are no duplicate \textit{variables} in the respective scope (i.e. either the global scope of the program or the local scope of the procedure), it is sufficient to check the corresponding \texttt{declarations}-attribute of \texttt{Program} and \texttt{Procedure} for any duplicates (see listing \ref{figure:duplicate-variables-contribution}).

\begin{minipage}{\linewidth}
\begin{lstlisting}[
  style=JastAdd,
  caption={Error contribution for duplicate \textit{variables}}, 
  label=figure:duplicate-variables-contribution
]
Program contributes "duplicate variables"
  when new HashSet<>(declarations()).size()
    != declarations().size()
  to Program.nameErrors();

Procedure contributes "duplicate variables"
  when new HashSet<>(declarations()).size() 
    != declarations().size()
  to Program.nameErrors();
\end{lstlisting}
\end{minipage}

\section{Type Analysis}

Type analysis, just like the name analysis, is divided into two parts: One that gathers the type information (either by synthesizing or inferring types) and one that checks if the program is well-typed. Clearly, the latter part may yield errors as a result, which again can be collected using collection attributes.

Similar to the approach for the name analysis, it is feasible to derive appropriate attribute equations from the formal specification of the type system. That is, given the set of type rules, attributes can be defined following a method. 

Usually, the aspect of a type system which handles the language's expressions is the most challenging one. This is primarily due to the fact that statements or definitions do not have a type at all. For that reason, we will restrict us here to the type rules for \ac{spl}'s expressions (see figure \ref{figure:spl-type-rules}). Moreover, since \ac{spl}'s type system is deliberately simple, types do not need to be inferred but only be synthesized.

\begin{mathpar}
  \inferrule[(Int)]{\;}{\Gamma \vdash \ell: int}
  \and
  \inferrule[(Arithmetic)]{\Gamma \vdash T_1: int \quad \Gamma \vdash T_2: int}{\Gamma \vdash T_1 +_{bin} T_2: int}
  \and
  \inferrule[(Comparison)]{\Gamma \vdash T_1: int \quad \Gamma \vdash T_2: int}{\Gamma \vdash T_1 \leq_{cmp} T_2: bool}
  \and
  \inferrule[(Negative)]{\Gamma \vdash T: int}{\Gamma \vdash -T: int}
  \and
  \inferrule[(Variable)]{\;}{\Gamma, x: \tau \vdash x: \tau}
  \and
  \inferrule[(Array Access)]{\Gamma \vdash T_i: array[\ell]\;of\;\tau \quad \Gamma \vdash T_j: int}{\Gamma \vdash T_i[T_j]: \tau}
\end{mathpar}
\captionof{figure}{The type rules for \ac{spl} expressions} 
\label{figure:spl-type-rules}
\vspace*{3mm}

Besides the scoping rules and type rules, it is necessary to specify in which case two types can be considered equal (e.g., to check the well-formedness of an assignment). The two most common type equivalences are \textit{name} and \textit{structural equivalence}. Yet, \ac{spl} uses neither of them. Instead, it uses an equivalence which we call \textit{referential equivalence} (as \ac{spl} introduces reference semantics for its types). Thanks to this special kind of type equivalence, Java's standard equality operator (\texttt{==}) can be used to test if two types are equal. In general, however, it is a much better idea to introduce a new method for types named \texttt{isEqualTo}, which implements the respective type equivalence. In the presence of subtyping it is conceivable to introduce another method \texttt{isSubTypeOf} as well.

To be able to reference the exact same type twice in a program, \ac{spl} provides type synonyms. The additional rule in figure \ref{figure:spl-type-subs} expresses the fact that a type synonym is equal to its referenced type. More technically it is necessary to resolve all type synonyms in JastAdd, as it would not be possible to compare two types using Java's equality operator otherwise. To illustrate this, consider for example comparing a named type \texttt{x} which is synonymous for another type $\tau$. Clearly, the Java objects representing the types \texttt{x} and $\tau$ are not the exact same object and thus do not share the same reference.

\begin{mathpar}
\inferrule[(Type Substitution)]{\Gamma,x = \tau \vdash T: x}{\Gamma, x = \tau \vdash T: \tau}
\end{mathpar}
\captionof{figure}{The type rule for type synonyms} 
\label{figure:spl-type-subs}
\vspace*{3mm}

Similar to the circumstance that no explicit symbol table is necessary when using \acp{rag}, no explicit data structure is required to model types. The subset of the \ac{ast} that represents type expressions can be reused to represent types during the type analysis. The two methods \texttt{isEqualTo} and \texttt{isSubTypeOf} may then be implemented as parameterized attributes. Moreover, type synonyms can be implemented by rewriting the \ac{ast}, which is very well supported by JastAdd or more generally rewritable \acp{rag}.

\begin{minipage}{\linewidth}
\begin{lstlisting}[
  style=JastAdd,
  caption={Type synonyms using rewritable \acp{rag}}, 
  label=figure:rewrite-type-synonyms
]
rewrite NameType {
  when (lookup(getName()).orElse(null) 
    instanceof TypeSynonym)
  to Type ((TypeSynonym) lookup(getName()).get())
    .getType();
}
\end{lstlisting}
\end{minipage}

Listing \ref{figure:rewrite-type-synonyms} shows the rewrite rule that implements the type rule of figure \ref{figure:spl-type-subs}. In fact, this rewrite rule may be even read in a similar way to the type rule: Given a type $x$ which is synonymous for $\tau$ (i.e. $x$ has to be a named type), we can also use type $\tau$ instead of $x$. More precisely, a named type (\texttt{NameType}) is rewritten if the condition after the keyword \texttt{when} holds. This condition stands for the premise that $x$ has to be a type synonym in the current context (which is expressed by the notation $\Gamma,x=\tau$ in the formal specification). The result of the rewriting process is the type for which the named type is synonymous.

\subsection{A Method for Deriving a Type Analysis}

As mentioned previously, type information only needs to be synthesized by inspecting the respective expression in \ac{spl} (as opposed to type inference). Although we are confident that it is generally conceivable to derive attribute equations from a type system that supports type inference (e.g. Hindley-Milner), we are still in progress to formalize our results to an extent where they lead to a pattern applicable to type inference.

Hence, we will restrict us to type synthesis in the rest of this paper. In general, synthesizing type rules can be translated to attribute equations using the following method: 

\begin{enumerate}
  \item Introduce a new synthesized attribute \texttt{type} for all non-terminals which formally have a type
  \item Introduce a new collection attribute \texttt{typeErrors} similar to \texttt{nameErrors}
  \item For each type rule that assigns a type $\tau$ to one of the non-terminals of step 1:
    \begin{enumerate}
      \item Add an equation that assigns $\tau$ to the \texttt{type}-attribute of the respective non-terminal.
      In case there are multiple type rules for the same syntactic construct conflicts may arise. These conflicts have to be resolved by either evaluating the context, by removing the conflicting type rules or by further distinguishing the syntactic constructs. 
      If $\tau$ depends on a premise (e.g. see the type rule \textsc{(Array Access)} where the result type depends on the left premise), a bottom type $\bot$ has to be returned if this premise is not fulfilled. This bottom type should be equal to any other type, which is most easily achieved by appropriately extending the parameterized attributes \texttt{isEqualTo} and \texttt{isSubTypeOf}.
      \item Add a contribution to the \texttt{typeErrors} attribute for each premise of the type rule
    \end{enumerate}
\end{enumerate}

Clearly, there are other premises (e.g. the condition of an \textit{if-statement} has to have a boolean type), which we have not shown here for the sake of brevity. Each of these requires a contribution to the \texttt{typeErrors} attribute similar to step 3b.

\subsection{An Exemplary Implementation of \textsc{(Int)} and \textsc{(Array Access)}}

To demonstrate the presented method, we will use it to derive the respective attributes and their equations for the type rules \textsc{(Int)} and \textsc{(Array Access)} presented earlier.

As shown in listing \ref{figure:type-and-errors-attribute}, two new attributes are required according to step 1 and 2. The first one is the \texttt{type}-attribute which is defined for all expressions (statements and declarations do not have a type in \ac{spl}). The collection attribute \texttt{typeErrors} is defined similarly to the \texttt{nameErrors}-attribute in subsection \ref{section:name-analysis}.

\vspace*{3mm}
\begin{minipage}{\linewidth}
\begin{lstlisting}[
  style=JastAdd,
  caption={The \texttt{type}-attribute for \ac{spl} expressions and the global \texttt{typeErrors}-attribute}, 
  label=figure:type-and-errors-attribute
]
syn Type Expression.type();
coll ArrayList<String> Program.typeErrors();
\end{lstlisting}
\end{minipage}

For the sake of brevity, we apply step 3 in this example only to the two type rules \textsc{(Int)} and \textsc{(Array Access)}. In general, a ``type rule that assigns a type $\tau$ to one of the non-terminals of step 1'' can be identified by the form of the conclusion. That is, if the conclusion depicts the form $\Gamma \vdash e : \tau$ it assigns $\tau$ to the non-terminal that represents the corresponding class of $e$. For example, in case of the type rule \textsc{(Int)}, $e$ is an integer literal $\ell$ and thus the type $int$ is assigned to the non-terminal we named \texttt{IntegerLiteral}. In fact, this already corresponds to the equation of step 3a (see listing \ref{figure:int-type-attribute}). Because the type rule \textsc{(Int)} is an axiom (i.e. it has no premise), step 3b can be skipped.

\vspace*{3mm}
\begin{minipage}{\linewidth}
\begin{lstlisting}[
  style=JastAdd,
  caption={The equation for the \texttt{type}-attribute of \textsc{(Int)}}, 
  label=figure:int-type-attribute
]
eq IntegerLiteral.type() = intType;
\end{lstlisting}
\end{minipage}

In contrast to \textsc{(Int)}, the type rule \textsc{(Array Access)} has two premises. But first things first: The conclusion of the type rule assigns each array access the base type $\tau$ of the accessed array. 
As defined in step 3a, when the result type depends on a premise, the bottom type $\bot$ has to be returned if this premise is not fulfilled. This is the case if the expression $T_i$ in the premise of rule \textsc{(Array Access)} is not an array (see listing \ref{figure:array-type-attribute}).

\begin{minipage}{\linewidth}
\begin{lstlisting}[
  style=JastAdd,
  caption={The equation for the \texttt{type}-attribute of \textsc{(Array Access)}}, 
  label=figure:array-type-attribute
]
eq Access.type() =
  getExp().type().isArrayType()) 
    ? ((ArrayType) getExp().type()).getBaseType()
    : bottomType;
\end{lstlisting}
\end{minipage}

As mentioned previously, the \textsc{(Array Access)} rule has two premises: One requiring that the accessed expression is an array and another one requiring that the index which is used to access the array is an integer. Although the implementation of the premises might be more complex depending on the actual language, the proceeding remains the same. By negating the premise, the error contribution can be written very naturally (as shown in listing \ref{figure:array-error-contributions}). The premises themselves can be implemented using a custom attribute (e.g. \texttt{expHasToBeAnArray()} and \texttt{indexHasToBeAnInteger()}).

\begin{lstlisting}[
  style=JastAdd,
  caption={The respective error contributions for the premises of the type rule \textsc{(Array Access)}}, 
  label=figure:array-error-contributions
]
Access contributes "illegal access"
  when !expHasToBeAnArray()
  to Program.typeErrors();

Access contributes "illegal index"
  when !indexHasToBeAnInteger()
  to Program.typeErrors();
\end{lstlisting}

Together, listing \ref{figure:array-type-attribute} and \ref{figure:array-error-contributions} implement the type rule \textsc{(Array Access)}. While the synthesized attribute constitutes the conclusion, the error contributions represents the premises.

\section{Summary}\label{Summary}
We have shown that a formal type system for a language can constructively and methodologically be translated into a specification for a \ac{rag}-based tool. The approach has been successfully implemented to generate a compiler (including code generation) for SPL that  performs (in terms of memory and time used for compilation) as well as a hand-crafted compiler.

The method is based on the JastAdd \ac{rag} mechanisms and does not require extensions. Furthermore, because these mechanisms are not exclusive to JastAdd but are general extensions to \acp{ag} and \acp{rag}, it is feasible to adapt our method to other tools as well. 
Even though SPL is a simple imperative language, further analysis has indicated that an extension of the approach is possible. For example, type inference according to Hindley-Milner should be implementable using JastAdd and the well-known Algorithm W \cite{milner-type-poly}. Because of the paradigm of \acp{ag} (i.e. attributes should not have externally visible side effects) the introduction of an appropriate monad (as described in \cite{journals/jfp/JonesVWS07}) could be necessary, though.
 
We are confident, that the approach to start not only with formal  grammars for lexical and syntax analysis, but also with a formal type system brings benefits to language designers since - as shown - this formalism can be translated into a specification for JastAdd (and potentially other tools), and thus eliminates the need for hand-written framework code in a compiler's name analysis and type analysis. An implementation of the method directly into JastAdd seems possible, but hasn't been started yet due to resource constraints.

\begin{acronym}
  \acro{ag}[AG]{Attribute Grammar}
  \acro{rag}[RAG]{Reference Attributed Grammar}
  \acro{cfg}[CFG]{Context-Free Grammar}
  \acro{spl}[SPL]{Simple Programming Language}
  \acro{ast}[AST]{Abstract Syntax Tree}
  \acro{thm}[THM]{Technische Hochschule Mittelhessen}
  \acro{nta}[NTA]{Non-Terminal attribute}
\end{acronym}

\setcitestyle{numbers,sort&compress}
\bibliography{types}{}
\bibliographystyle{ACM-Reference-Format}
\end{document}